# Superconductivity at 17 K in (Fe$_2$P$_2$)(Sr$_4$Sc$_2$O$_6$): a new superconducting layered pnictide oxide with a thick perovskite oxide layer

Hiraku Ogino[1,2], Yutaka Matsumura[1], Yukari Katsura[1], Koichi Ushiyama[1], Shigeru Horii[1,2], Kohji Kishio[1,2] and Jun-ichi Shimoyama[1,2]

[1]Department of Applied Chemistry, The University of Tokyo, 7-3-1 Hongo, Bunkyo-ku, Tokyo 113-8656, Japan
[2]JST-TRIP, Sanban-cho, Chiyoda-ku, Tokyo 102-0075, Japan
e-mail address: tuogino@mail.ecc.u-tokyo.ac.jp

## Abstract

A new layered oxypnictide (Fe$_2$P$_2$)(Sr$_4$Sc$_2$O$_6$) have been synthesized by solid-state reaction. This material has an alternating layer stacking structure of anti-fluorite Fe$_2$P$_2$ and perovskite-based Sr$_4$Sc$_2$O$_6$ oxide layers. Space group of the material is *P4/nmm* and lattice constants *a* and *c* are 4.016 Å and 15.543 Å, respectively. The interlayer Fe-Fe distance corresponding to the *c*-axis length is the longest ever reported in the iron-based oxypnictide systems. In both magnetization and resistivity measurements, the present compound exhibited superconductivity below 17 K, which is much higher than that of LaFePO and the highest in arsenic-free iron-based oxypnictide systems under ambient pressure.

## Introduction

Since the discovery of high-$T_c$ in LaFeAs(O,F)[1], new superconductors containing anti-fluorite iron pnictide layer have been eagerly searched and several group of superconducting materials, such as LiFeAs[2], *AE*Fe$_2$As$_2$(*AE* = Alkali earth metals)[3], *RE*FeAsO(abbreviated as 1111, *RE* = Rare earth elements)[4], *AE*FeAsF[5] and related oxyphosphide and chalcogenide materials have been discovered thus far.

Empirically, the high $T_c$ exceeding 50 K was achieved in the fluorine-doped or oxygen deficient REFeAsO [6-8] or RE doped *AE*FeAsF systems[9]. Crystal structures of these materials can be regarded as layer stacking of anti-fluorite-type FeAs layer and fluorite-type oxide or fluoride layer. Several discussions have been already made for determining factors of $T_c$ for iron-based oxypnictides. From a structural viewpoint of superconducting iron oxypnictide layer, Lee *et al*. has pointed out that high symmetry in FeAs$_4$ tetrahedra is crucial for high $T_c$[10]. Actually, the highest $T_c$ for 1111 and 122 phases were achieved by compounds with angles between As-Fe-As of ~109.5°, which is the desirable value for perfect symmetry. Based on this idea, relatively low $T_c$'s in LaFeP(O,F) and FeSe are reasonably explained.

On the other hand, materials with similar stacking structure of perovskite-type oxide layer and anti-fluorite chalcogenide or pnictide layer have been already found in several oxypnictides[11] and oxychalcogenide systems[12]. Because of flexibility in the perovskite-based structure, several structure types and variety of constituent transition metal in the perovskite block have been reported[12-18]. In our previous study[14], several new oxysulfides including a scandium-based oxysulfide $(Cu_2S_2)(Sr_3Sc_2O_5)$ (abbreviated as Sc-22325) were discovered. Recently isostructural compound having FeAs layer was reported by Zhu *et al*[19], while it did not show superconductivity probably due to insufficient carrier concentration. These facts indicate structural similarity between layered iron oxypnictides and layered cupper oxysulfides. Because of structural and chemical variety of the cupper oxysulfides, oxypnictides with perovskite oxide layer is expected to be repository of new functional materials including superconductors.

In the present study, iron oxypnictide with new structure $(Fe_2P_2)(Sr_4Sc_2O_6)$ (abbreviated as Sc-22426) have been successfully synthesized. This material has perovskite-related $Sr_4Sc_2O_6$ layer($K_2NiF_4$-type), which results in the longest interlayer Fe-Fe distance ~15.5 Å, and showed bulk superconductivity with $T_c$(onset) of 17 K.

**Experimental**

All samples were synthesized by solid-state reaction starting from FeP, SrO(2N), Sr(2N), $Sc_2O_3$(4N). The precursor FeP was obtained by the chemical reaction of Fe(3N) and P(2N) with a molar ratio of 1:1 at 700°C for 12 h. The nominal composition was fixed according to the general formula : $(Fe_2P_2)(Sr_4Sc_2O_6)$. Since the starting reagents, Sr and SrO, are sensitive to moisture in air, manipulations were carried out under an inert gas atmosphere. Powder mixture of FeP, SrO, Sr and $Sc_2O_3$ was pelletized and sealed in evacuated quartz ampoules. Heat-treatments were performed in the temperature range from 900 to 1200°C for 12 to 40 hours.

Phase identification was carried out by X-ray diffraction with RIGAKU Ultima-IV diffractmeter and intensity data were collected in the $2\theta$ range of 5° - 80° at a step of 0.02°(Cu-$K_\alpha$) and Si was used as internal standard. Structural refinement was performed using the analysis program RIETAN-2000[20]. High-resolution images were taken by JEOL JEM-2010F field emission TEM. Magnetic susceptibility measurement was performed by a SQUID magnetometer (Quantum Design MPMS-XL5s). Electric resistivity was measured by AC four-point-probe method using Quantum Design PPMS under fields up to 9 T.

**Results and Discussions**

Samples containing Sc-22426 as a main phase were obtained by sintering above 1100°C for longer than 24 h. The color of samples was dark black, however, they expanded during sintering, resulting in low bulk density less than 40%. This is possibly due to the anisotropic grain growth of Sc-22426 crystals or intermediate products during ramping up process before sintering process..

Figure 1(a) shows calculated and observed XRD patterns of Sc-22426 reacted at 1200°C for 40 h. This compound consist of stacking of antifluorite $Fe_2P_2$ layer and perovskite type $Sr_4Sc_2O_6$ layer as presented in Fig. 1(b). Sc-22426 was obtained as the main phase by optimization the sintering condition, while formation of the secondary phase $SrFe_2P_2$ is inevitable at the present stage. All peaks were indexed by Sc-22426 or $SrFe_2P_2$. The ratio of Sc-22426 : $SrFe_2P_2$ was estimated to be ~ 9 : 1. Space group of the Sc-22426 was *P4/nmm* and its lattice constants were determined to be $a$ = 4.016 Å and $c$ = 15.543 Å. The interlayer Fe-Fe distance of the material is extremely long compared to 8.5 Å in LaFePO and even longer than recently reported 13.4 Å in Sc-22325 with FeAs layer[19]. Although Rietveld refinement was not well fitted probably co-existence of unidentified impurities, Fe-P-Fe angles calculated using tentative value of the coodinates were 74.8° and 118°, which is closer to ideal angle of highly symmetric tetrahedra compared to that of LaFePO[8]. The Fe-Fe distance of 2.840 Å is also close to that of LaFeAsO rather than LaFePO. These facts suggest that the new system has different nature of local iron pnictide structure from that in the 1111 system.  In addition, it would be possible to optimize local structure of pnictide layer by changing the size of the perovskite subunit. Figure 2 shows a bright-field TEM image and an electron diffraction pattern taken from [1 -1 0] direction of a Sc-22426 crystal. Both TEM image and electron diffraction patterns indicated tetragonal cell with $a$ ~ 4.0 Å and $c$ ~ 16 Å, which coincide well with corresponding values obtained from XRD patterns.  It should be noted that any stacking faults were not found in the observed crystals, whereas the perovskite block has large variety in layer stacking pattern.  Furthermore, absence of satellite spots probes commensurate stacking between $Fe_2P_2$ and $Sr_4Sc_2O_6$ layers.

Temperature dependences of zero-field-cooled(ZFC) and field-cooled(FC) magnetization of Sc-22426 measured under 1 Oe are shown in Fig. 3. As clearly seen, Sc-22426 showed large diamagnetism due to superconductivity with $T_{c(onset)}$ of ~17 K. Superconducting volume fraction estimated from ZFC magnetization at 2 K was much larger than the perfect diamagnetism.  This extremely large diamagnetism is due to the porous microstructure having large amount of closed pore and demagnetization effect of the sample.  The reversible region, where ZFC and FC magnetization curve overlaps each other, down to ~12 K suggested that both grain coupling and intragrain pinning are relatively poor.

Figure 4 shows temperature dependence of resistivity for Sc-22426. The resistivity under 0 T was measured up to 300 K as shown in the inset. Although the metallic behavior was observed in the normal state resistivity, absolute value of $\rho$ is approximately two orders of magnitude higher than that reported for 1111 compounds. Superconducting transition was observed at $T_{c(onset)}$ of 17 K, while zero resistivity was achieved at 9.8 K.  Under magnetic fields, the superconducting transition became extremely broad and determination of $T_{c(onset)}$ to estimate upper critical fields was difficult.  The observed high resistivity and broad superconducting transition under magnetic fields are explained by porous microstructure and poor grain connectivity of the sample. In addition, high electromagnetic anisotropy due to thick perovskite layer, *i.e.,* blocking layer might contribute broad resistivity transition in fields.

The newly found Sc-22426 with FeP layer shows clear superconductivity below 17 K. This $T_c$ is much higher than that of LaFeP(O,F) and even higher than FeSe under ambient pressure[21,22]. To our knowledge, the $T_c$ of the present compound is the highest among the arsenic-free iron-based systems under ambient pressure. Its high $T_c$ might partially attributed to local structure of FeP layer having relatively high symmetry. Further investigation will be needed to clarify the essential effects of very long Fe-Fe interlayer distance on electronic and magnetic nature of the present system. Another advantage of the new material is large variations of doping schemes and perovskite-type structure. Since the perovskite-type $Sr_4Sc_2O_6$ layer is structurally and chemically flexible, control of carrier concentration by doping for all sites by various elements will be possible. In addition, complete substitution of $Sr^{2+}$ and $Sc^{3+}$ sites and change of the perovskite structure to control thickness of blocking layer and *a*-axis length will be promising to find new superconductors having higher $T_c$. In other words, feasibility of doping or structural optimization in the perovskite-layer containing compounds has more options than in the 1111 system.

**Conclusions**

A new layered iron oxypnictide Sc-22426; $(Fe_2P_2)(Sr_4Sc_2O_6)$ with the space group of *P4/nmm* was successfully synthesized. This material has layer stacking of anti-fluorite $Fe_2P_2$ and perovskite $Sr_4Sc_2O_6$ layers and its interlayer Fe-Fe distance of 15.5 Å is the longest among the reported iron oxypnictides. In both magnetization and resistivity measurements, superconducting transition was observed with $T_{c(onset)}$ of ~17 K. This high $T_c$ is partly due to the higher tetrahedral symmetry at FeP layer than in other iron oxyphosphides. Our results strongly indicate that oxypnictides having perovskite oxide layer can be a "new family" of iron-based superconductors and the large variety of perovskite-type structures and constituent elements will open new windows for search of oxypnictide superconductors.

**Ackowledgement**

This work was partly supported by Center for Nano Lithography & Analysis, The University of Tokyo, supported by the Ministry of Education, Culture, Science and Technology (MEXT), Japan.

**Figure captions**

Figure 1. Powder XRD pattern of the Sc-22426 (a), calculated XRD pattern of mixture of Sc-22426 : $SrFe_2P_2$ = 9:1 (b) and crystal structure of Sc-22426 (c). Stars indicate the positions of diffraction peaks due to a secondary phase $SrFe_2P_2$.

Figure 2. Bright-field TEM images and corresponding electron diffraction patterns of a Sc-22426 crystal viewed from [1 -1 0] direction

Figure 3. Temperature dependence of ZFC and FC magnetization curves of the Sc-22426 bulk measured in 1 Oe.

Figure 4. Temperature dependences of resistivity of the Sc-22426 bulk under magnetic fields from 0 T to 9 T.   Temperature dependence of resistivity from 2 K to 300 K in 0 T is shown in the inset.

Figure 1

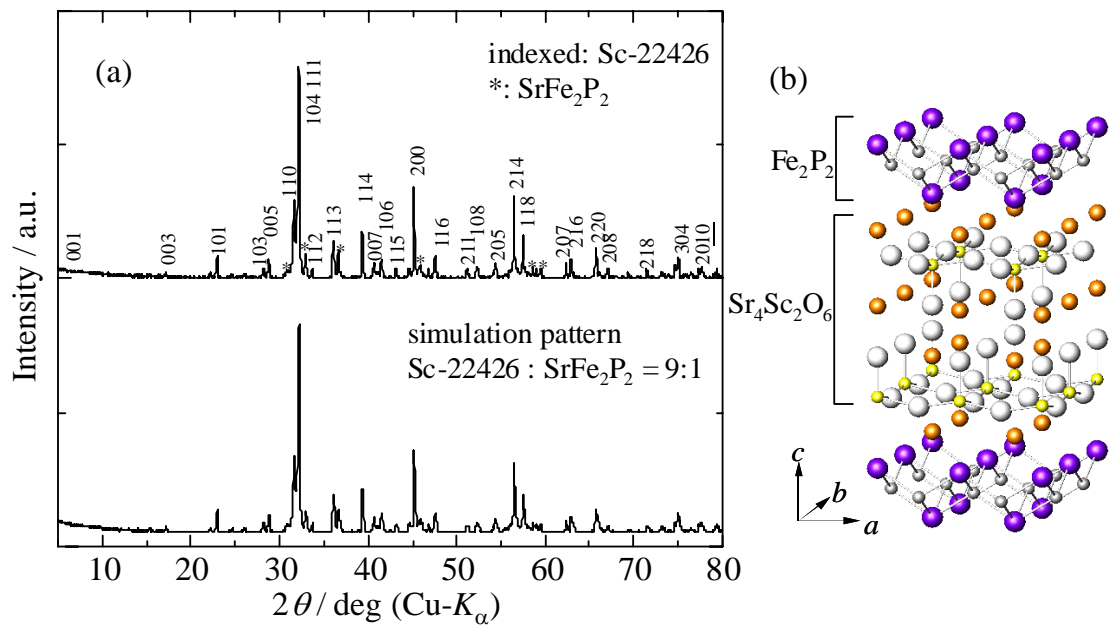

Figure 2

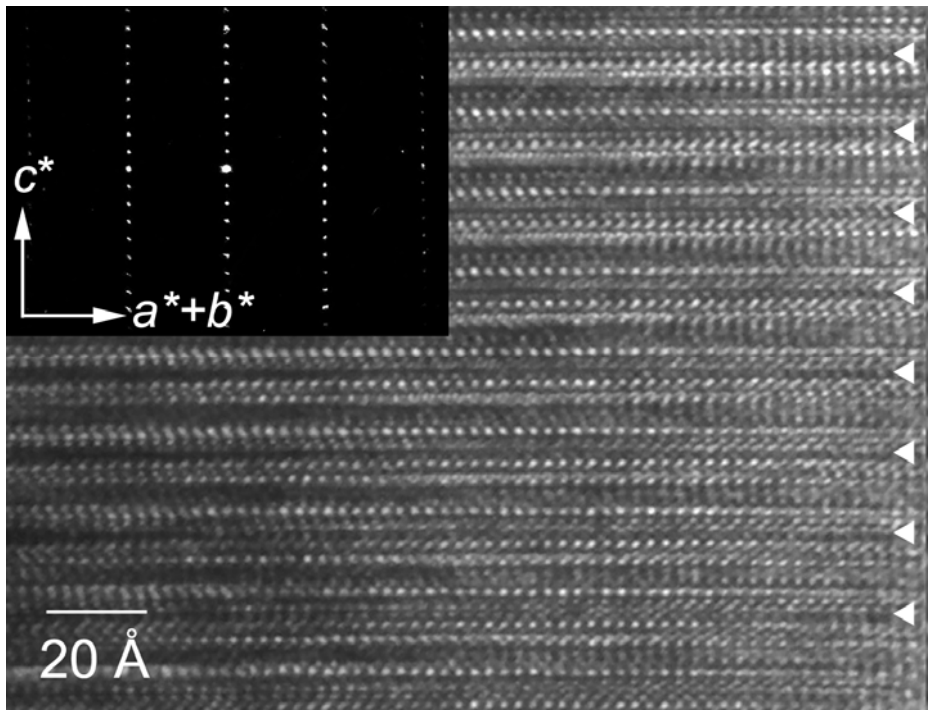

Figure 3

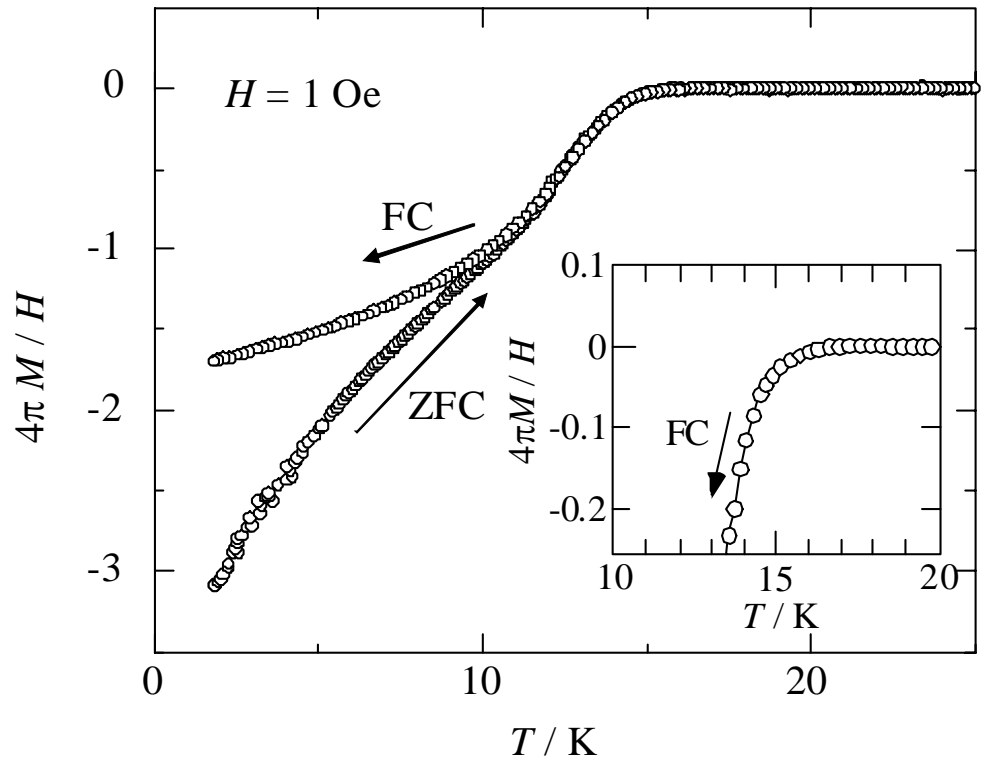

Figure 4